\documentclass[a4paper]{article}

\usepackage{INTERSPEECH2022}
\usepackage{siunitx}
\usepackage{multirow}
\usepackage{makecell}
\usepackage{url}
\usepackage{todonotes}
\usepackage{caption}
\title{Unify and Conquer: How Phonetic Feature Representation Affects Polyglot Text-To-Speech (TTS)}
\name{Ariadna Sanchez, Alessio Falai, Ziyao Zhang,  Orazio Angelini, Kayoko Yanagisawa}
\address{
  Alexa AI, Amazon
}
\email{\{cariadn,falai,zhaziyao,aorazio,yakayoko\}@amazon.com}

\begin{document}

\maketitle
\begin{abstract}
An essential design decision for multilingual Neural Text-To-Speech (NTTS) systems is how to represent input linguistic features within the model. Looking at the wide variety of approaches in the literature, two main paradigms emerge, unified and separate representations. The former uses a shared set of phonetic tokens across languages, whereas the latter uses unique phonetic tokens for each language. In this paper, we conduct a comprehensive study comparing multilingual NTTS systems models trained with both representations. Our results reveal that the unified approach consistently achieves better cross-lingual synthesis with respect to both naturalness and accent. Separate representations tend to have an order of magnitude more tokens than unified ones, which may affect model capacity. For this reason, we carry out an ablation study to understand the interaction of the representation type with the size of the token embedding. We find that the difference between the two paradigms only emerges above a certain threshold embedding size. This study provides strong evidence that unified representations should be the preferred paradigm when building multilingual NTTS systems.
\end{abstract}

\section{Introduction}
\label{sec:introduction}
Text-To-Speech (TTS) has been attempting to move from monolingual to multilingual applications for a long time \cite{javkin1989multilingual}. Advancements in Neural TTS (NTTS) now allow us to create high-quality multilingual and multi-speaker speech synthesis models \cite{zhang2019learning,do2021systematic}. Multilingual TTS models differ from each other significantly in terms of the level of flexibility of speaker/language combinations that they are able to deliver. Some models rely on bi-/multi-lingual recordings from the same speaker to generate multilingual synthesis \cite{ming2017light,maiti2020generating}, i.e. speaker and language are coupled. Other multilingual models are capable of generating speech in languages for which there are no available recordings from the speaker \cite{latorre21_ssw}. In this paper, we give the name Polyglot TTS to the multi-speaker multilingual TTS system that enables any speaker in the training corpus to speak any language present in it.

From a linguistic and design perspective, modelling input features for Polyglot TTS systems presents an interesting research question. Previous Polyglot TTS work has used various input representations. We divide them into two main groups: \textit{unified representations}, which are normally based on the International Phonetic Alphabet (IPA) \cite{international1999handbook}, and \textit{separate representations}, which entail having separate input tokens for each language.
For \textit{unified representations}, there is previous work based on the IPA features \cite{hemati2020using, zhang2021revisiting}. Other works used phonetic features extracted from dictionaries like CMUDict \cite{cai2020cross}, or using a handcrafted phone-set \cite{demirsahin2018unified}, which is not a scalable approach. Some previous work also defines phones with phonological features instead of using a single token per phoneme, thus allowing for zero-shot multilingual synthesis \cite{staib2020phonological,maniati2021cross}. There are other unified representation approaches which, instead, have used text-based features such as characters \cite{prakash19_ssw} or bytes \cite{li2019bytes, he2021multilingual}. 

For \textit{separate representations}, \cite{zhang2019learning} studies the use of graphemes, byte-encoding and phonemes by concatenating symbols from all the language vocabularies, with common tokens shared between them. Another work \cite{tu2019end} investigates learning phonetic mappings between two languages. In \cite{nachmani2019unsupervised}, authors use the same phonetic representation but separate phoneme encoders for each language.

Conceptually, there are arguments in favour of both paradigms. For example, having separate tokens per language and accent may help the model avoid accent leakage, i.e. the synthesis having a slight accent from the original language of the target speaker, because all phonetic tokens are unique. On the other hand, we can also hypothesise that a language-agnostic solution can improve the accent and naturalness by removing redundant tokens and thus allowing the model to generalise over similarities between languages and accents.

The main contribution of this paper lies in the experimentation and evaluation aimed at understanding how \textit{unified representations} and \textit{separate representations} of input linguistic features affect the quality of Polyglot synthesis in terms of naturalness and accent of the voice. 
To the best of our knowledge, this is the first work conducting a systematic study and evaluation on this subject.

\section{Methodology}
\label{sec:methodology}

\subsection{Polyglot Text-to-Speech}
\label{subsec:polyglot-tts}
\subsubsection{Architecture}
\label{subsubsec:architecture}
Our base architecture consists of a neural acoustic model and vocoder pair. The acoustic model, tasked to convert phonetic sequences into mel-spectrograms, follows \cite{latorre2019effect}, which is a sequence-to-sequence attention-based model \cite{sequence-to-sequence}. We use a universal vocoder based on \cite{jiao2021universal}, which takes as input mel-spectrograms and outputs raw waveforms.

\begin{figure}[t]
  \centering
  \includegraphics[width=\linewidth]{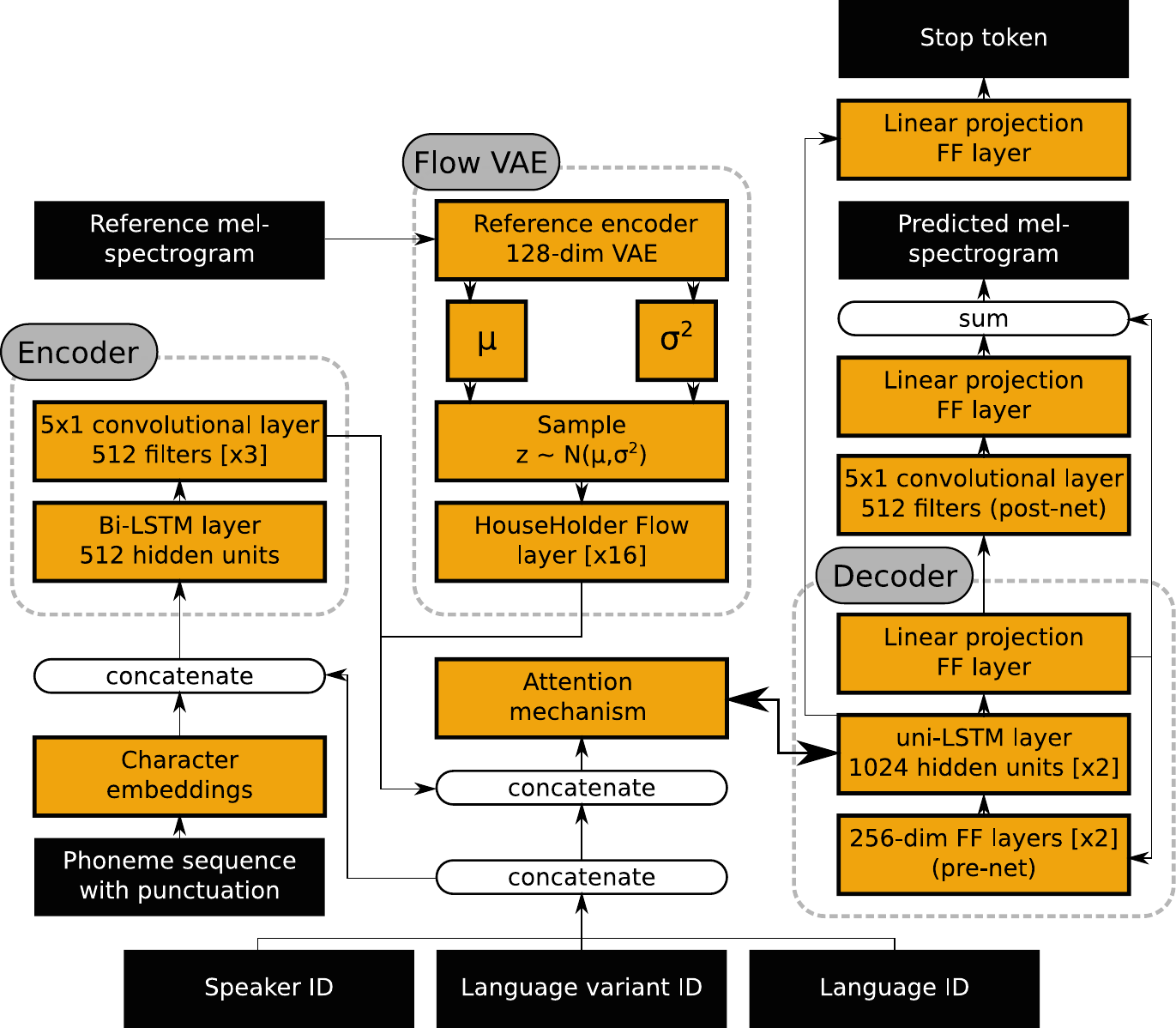}
  \caption{Architecture used for the acoustic model, explained in~Section \ref{subsubsec:architecture}. For some models, language conditioning is not used, as explained in Section~\ref{sec:experiments}.}
  \label{fig:arch}
\end{figure}

The acoustic model is composed of $3$ major components: an encoder, a decoder and an attention mechanism. Phonemes are given as input to the encoder, which processes them through a stack of $3$ convolutional blocks, followed by a single bi-LSTM unit. The decoder comprises a pre-net and a stack of $2$ LSTM layers, to then branch out into mel-spectrogram and stop-token heads, in order to predict the mel-bands for the current time-frames and the probability of cutting off auto-regressive generation, respectively. A final post-net acts as a residual signal to adjust mel-spectrogram predictions. In order to learn how many frames correspond to each input phoneme, we rely on a location-sensitive cell based on \cite{location-sensitive-attention}, with the same settings as reported in \cite{tacotron2}.

Multi-speaker, cross-lingual synthesis is enabled with lookup tables pointing at $256$-dimensional trainable embeddings to condition the encoder on speaker, language and language variant (LV) (refer to Table \ref{tab:locale_list}). These embeddings are fed to the encoder and the attention mechanism, as shown in Figure \ref{fig:arch}. To condition the model on \emph{style}\footnote{In this work, we define \emph{style} as any component of prosody that cannot be inferred solely from text.}, we rely on a reference encoder \cite{reference-encoder} composed of a Variational Auto-Encoder (VAE) followed by $16$ \emph{householder} flow layers \cite{householder-flow}. All embeddings are upsampled and concatenated to both the inputs and the outputs of the linguistic encoder.

At inference time, we are able to synthesise speech with the desired speaker, language, LV and \emph{style}, by conditioning the model on the appropriate embeddings. For speaker, language and LV, we simply select representations from each lookup table, while for \emph{style} we use a pre-calculated centroid of the available ground truth data for the target speaker in the target style, in place of the VAE-generated embedding.

All models are trained with the Adam optimiser and a combination of $L_1$, Kullback–Leibler Divergence (KLD) and Cross Entropy (CE) loss functions for spectrogram, VAE and stop token, respectively, similarly to \cite{shannon}. We also make use of an exponentially decaying learning rate, that starts at $10^{-3}$ and is reduced by 2\% every \SI{10}{k} steps, and follow the standard teacher forcing \cite{teacher-forcing} approach to predict $5$ mel-spectrogram frames at each iteration. Hyper-parameters (such as the learning rate schedule) are selected following standard practices in the literature \cite{tacotron, tacotron2}. Their particular values marginally affect the final results, but not the conclusions.

\subsubsection{Data composition}
\label{subsubsec:data-composition}
Our dataset consists of $130$ speakers from $12$ LVs across $4$ languages, which are summarised in Table~\ref{tab:locale_list}. LVs belong to either the Romance or Germanic language family. The dataset contains between $3$ and $6$ hours of recordings per speaker. The \emph{target speaker}, i.e. the speaker for which we want to enable cross-variant synthesis, has instead more than \SI{30}{h} of recordings. 
We do not impose a uniform balance between the amount of data in each language and LV, as it is beyond the scope of this paper. Overall, the dataset is composed of \SI{580}{h} of recordings.

\begin{table}[th]
    \centering
    \caption{\label{tab:locale_list} Language, language variant (LV) and code for all speakers included in our dataset.}
    \begin{tabular}{@{}ccc@{}}
    \toprule
    \textbf{Language} & \textbf{Language variant (LV)} & \textbf{Language variant code} \\ \midrule
    \multirow{5}{*}{English} & Australia & en-AU \\
     & India & en-IN \\
     & Wales & en-WLS \\
     & Britain & en-GB \\
     & US & en-US \\ \hline
    \multirow{2}{*}{French} & Canada & fr-CA \\
     & France & fr-FR \\ \hline
    \multirow{3}{*}{Spanish} & US & es-US \\
     & Spain & es-ES \\
     & Mexico & es-MX \\ \hline
    German & Germany & de-DE \\ \bottomrule
    \end{tabular}
\end{table}

\vspace{-0.1cm}
\subsection{Phonetic representations}
\label{subsec:phonetic-representations}
This paper compares two different input representations, which we call (language-)\textit{separate} and (language-)\textit{unified}, and are defined as follows.
\begin{itemize}
\itemsep0em 
\item \textit{Separate}: In this paradigm, there is a separate phone-set for each LV. In each phone-set, each phoneme has its own unique token. In other words, even phones which share the same symbol in different LVs have separate tokens in the input representation (e.g. \texttt{/b/} in American and British English or \texttt{/a/} in Canadian French and German).
\item \textit{Unified}: We use a unified representation of phones for all LVs in the training data. In this paradigm, the phones are denoted using X-SAMPA \cite{wells1995xsampa}, a machine-readable notation for the IPA. The same set of tokens is used across all LVs, e.g. \texttt{/b/} in American English and British English correspond to the same token.
\end{itemize}

In \textit{unified} features, both service tokens (such as word boundary and utterance start tokens) and punctuation tokens are shared across languages, just like the phoneme tokens. On the other hand, \textit{separate} representation models are trained with shared service tokens, but punctuation tokens are separate for each language, like the phoneme tokens. The reason for this is explained in Section~\ref{subsec:phonetic-representations-comparison}.

\section{Experiments}
\label{sec:experiments}
All subjective evaluations conducted are controlled in terms of training data compositions and the number of training epochs.
A slightly different architecture is used for the $2$ input paradigms. For \textit{separate} features, we condition the model on speaker, language and LV; while for \textit{unified} features, we only use speaker and LV conditioning. Subjective evaluations revealed that, when using \textit{separate} features, conditioning on speaker, language and LV is significantly better than conditioning only on LV. Contrarily, for \textit{unified} features, conditioning only on speaker and LV leads to significantly better accent, while also adding language conditioning brings accent degradation caused by the target speaker's own accent. We hypothesise that, for \textit{unified} features, when the model sees both language and LV conditioning, it diminishes their importance. This, in turn, makes the model more reliant on the speaker embedding, which contains language information of the target speaker, hence impacting the accent negatively. 

\subsection{Comparison of phonetic representations}
\label{subsec:phonetic-representations-comparison}
We compare \textit{separate} and \textit{unified} phonetic representations by training two models, one for each kind of input phonetic features. For both models, the target LVs we evaluate are de-DE, fr-CA, en-GB and es-US. 

We compare each model for naturalness and accent separately, using a MUltiple Stimuli with Hidden Reference and Anchor (MUSHRA) test \cite{recommendation2001bs}. 
For each of the target LVs, we synthesise $100$ samples from both models for evaluation. A total of $60$ listeners participate in each evaluation, who listen to $50$ randomly selected utterances from the pool of $100$ samples. 
For preserving the integrity of our evaluations, we use upper and/or lower anchors as control samples. We filter out any listeners that consistently give undue high/low scores to the lower/upper anchors, respectively. No more than $6$ listeners were excluded from each evaluation.

\begin{figure}[t]
  \centering
  \includegraphics[width=\linewidth]{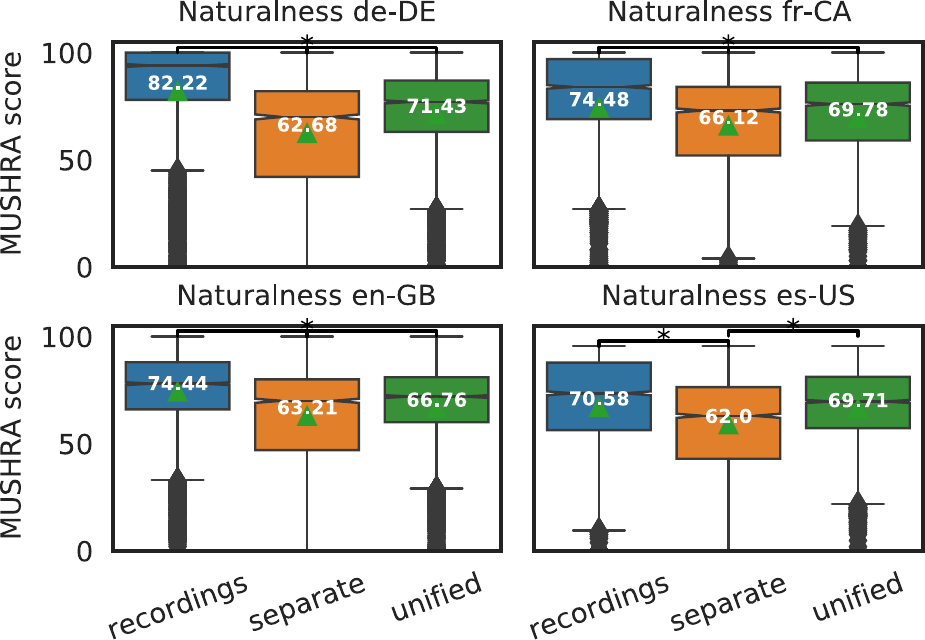}
  \caption{Naturalness MUSHRA results for de-DE, en-GB, fr-CA and es-US. See Section \ref{subsec:phonetic-representations-comparison} for explanation.}
  \label{fig:naturalness_degbfr}
\end{figure}

The systems included in naturalness evaluation are (1) \textit{separate} model, (2) \textit{unified} model, and (3) the original en-US target speaker recordings, which serve as the upper anchor. We ask the listeners to rate how natural each sample sounds on a scale of 0 (``completely unnatural'') to 100 (``completely natural''), ignoring any differences in language or accent. 
Figure \ref{fig:naturalness_degbfr} shows the results for each LV, where synthesis with \textit{unified} features scores higher than with \textit{separate} features, with p-value $\ll 0.001$ for all comparisons.

\begin{figure}[t]
  \centering
  \includegraphics[width=\linewidth]{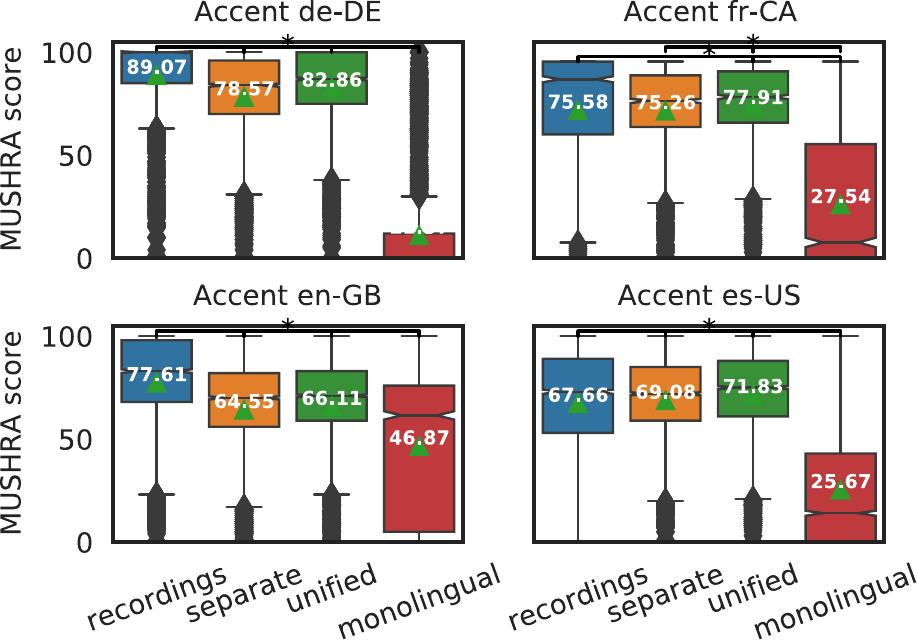}
  \caption{Accent MUSHRA results for de-DE, en-GB, fr-CA and es-US. See Section \ref{subsec:phonetic-representations-comparison} for explanation.}
  \label{fig:accent_degbfr}
\end{figure}

The systems included in accent evaluation are (1) \textit{separate} model, (2) \textit{unified} model, (3) the recordings from a native speaker of the target language as upper anchor, and (4) the synthesis from a model that has been trained with monolingual data of the target speaker (an American English speaker) as lower anchor. For the lower anchor, we map each phone of the target language with a similar phone that exists in the monolingual model, purposely producing American English-accented speech. We ask each listener to rate how native does each sample sound from 0 (``completely foreign'') to 100 (``completely native''). 
Figure \ref{fig:accent_degbfr} shows the MUSHRA results for each LV. Synthesis with \textit{unified} features scores higher than with \textit{separate} features in each case, with p-value $\ll 0.001$ for all comparisons, although the gaps between the mean values are smaller. 

\begin{figure}[t]
  \centering
  \includegraphics[width=\linewidth]{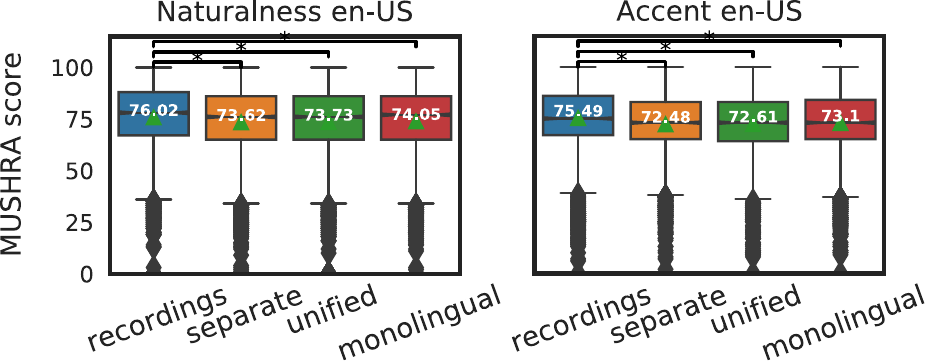}
  \caption{Naturalness and Accent MUSHRA results for en-US. See Section \ref{subsec:phonetic-representations-comparison} for explanation.}
  \label{fig:enus_plots}
\end{figure}

We have typically observed that, when a model is trained with multilingual corpora, there is quality degradation in the target speaker's synthesis in their native LV, which may be due to training data composition \cite{mix-n-match}. Therefore, we also evaluate the models in terms of naturalness, accent and speaker similarity for the native LV of the target speaker, i.e., en-US. Naturalness and accent are evaluated via MUSHRA tests. Following \cite{yourtts}, we use the pre-trained speaker verification model Resemblyzer\footnote{https://github.com/resemble-ai/Resemblyzer} to measure speaker similarity.
Figure \ref{fig:enus_plots} shows that the difference in both accent and naturalness between our proposed models and the monolingual model is not statistically significant. This suggests that both input features result in audio samples of comparable quality to that of the monolingual model when speaking the target speaker's native LV. 

Table \ref{tab:cosine_distance} shows the average cosine similarity between the samples used in the MUSHRA evaluations and the centroid of the original speaker. We observe that the monolingual model yields the best similarity against the recordings centroid, followed by the \textit{unified} representation model, and then the \textit{separate} representation model. A two-sided t-test shows that all differences are statistically significant (p-value $\ll 0.001$).

\begin{table}[th]
  \caption{Cosine similarity for speaker embeddings extracted with Resemblyzer. We report mean and standard deviation of the samples used in evaluations wrt the recordings centroid, calculated with \SI{5}{k} original recordings.}
  \label{tab:cosine_distance}
  \centering
  \begin{tabular}{ll}
    \toprule
    \textbf{Model} & \textbf{Cosine similarity} \\ \midrule
    \textit{recordings} & \num{0.8058958} $\pm$ \num{0.05893049} \\
    \textit{monolingual} & \num{0.7889244} $\pm$ \num{0.037844058} \\
    \textit{unified} & \num{0.76749384} $\pm$ \num{0.052282963} \\
    \textit{separate} & \num{0.7252414} $\pm$ \num{0.051151365} \\
    \bottomrule
  \end{tabular}
\end{table}

We additionally observe that the \textit{unified} representation model is more stable than the \textit{separate} representation model, with fewer occurrences of mumbling or skipped punctuation pauses. We hypothesise that reducing the number of phonetic tokens correspondingly reduces the number of contexts the model has to learn, while keeping the amount of data constant, i.e., the model sees more examples of each possible context and is poised to generalise better. This enables the model to learn a more robust representation of the phonetic features, leading to a more stable model. Sharing the punctuation tokens across LVs also helps to increase the amount of seen data for each punctuation, leading to better pausing. In the case of \textit{separate} features, merging the punctuation tokens across languages led to instabilities. We suspect that this is because for a given punctuation token (e.g. comma), there is around one order of magnitude more possible surrounding phonetic contexts compared to the \textit{unified} case, since there are more phonetic tokens. This, in effect, reduces the model's ratio of observed versus possible contexts, leading to instabilities. Hence for the \textit{separate} case we kept the punctuation tokens separate for each LV.

\subsection{Ablation study for phoneme embedding size}
\label{subsec:ablation-embedding-size}
The number of input tokens in the \textit{separate} and \textit{unified} paradigms differs significantly. We therefore investigate the effect of the size of the phoneme embedding vector in both cases.
In this section we describe an ablation study where the \emph{size of the phoneme embedding} $s$ is gradually reduced. This would confirm whether the results in Section \ref{subsec:phonetic-representations-comparison} still hold when the phoneme embedding size changes. It also reveals the minimum $s$ to avoid synthesis quality degradation. 

We train Polyglot models with \textit{separate} and \textit{unified} features using $s=64$ and $s=128$, while keeping all other parameters fixed. We then compare such reduced capacity models between themselves and against the ones reported in Section \ref{subsec:phonetic-representations-comparison}, for which $s=256$.

Specifically, we run $2$ different sets of MUSHRA tests for naturalness and accent: one comparing \textit{separate} and \textit{unified} features with each other at lower embedding sizes (results in Table \ref{tab:ablation-lower-comp}); and another to identify the best $s$ parameter for both \textit{separate} and \textit{unified} features (results in Table \ref{tab:ablation-best-size}). \vspace{-0.25cm}

\begin{table}[th]
    \centering
    \setlength\tabcolsep{3pt}
    \renewcommand{\arraystretch}{1.2}
    \caption{Mean MUSHRA scores relative to recordings for naturalness (N) and accent (A), comparing different phoneme embedding sizes within the \textit{separate} and \textit{unified} features, across all evaluated LVs.}
    \label{tab:ablation-best-size}
    \begin{tabular}{@{}ll|lll|lll@{}}
    \toprule
     &  & \multicolumn{3}{c|}{\textbf{Separate}} & \multicolumn{3}{c}{\textbf{Unified}} \\
     &  & \multicolumn{1}{c}{$64$} & \multicolumn{1}{c}{$128$} & \multicolumn{1}{c|}{$256$} & \multicolumn{1}{c}{$64$} & \multicolumn{1}{c}{$128$} & \multicolumn{1}{c}{$256$} \\  \midrule
    \multirow{4}{*}{\textbf{N}} & \textbf{en-GB} & \textbf{\num{83.40}} & 
    \num{82.79} & \num{81.72} & \num{78.70} & \num{74.99} & \textbf{\num{85.01}} \\
     & \textbf{fr-CA} & \num{92.15} & \num{87.65} & \textbf{\num{95.01}} & \num{89.25} & \num{90.29} & \textbf{\num{98.67}} \\
     & \textbf{de-DE} & \num{78.00} & \textbf{\num{79.65}} & \num{77.81} & \num{85.66} & \num{83.76} & \textbf{\num{91.46}} \\
     & \textbf{es-US} & \num{94.12} & \textbf{\num{95.22}} & \num{92.79} & \num{95.92} & \num{99.33} & \textbf{\num{107.91}} \\ \midrule 
    \multirow{4}{*}{\textbf{A}} & \textbf{en-GB} & \textbf{\num{82.97}} & \num{82.94} & \num{81.83} & \num{84.17} & \num{82.83} & \textbf{\num{85.90}} \\
     & \textbf{fr-CA} & \num{92.44} & \num{89.19} & \textbf{\num{93.60}} & \num{90.91} & \num{92.57} & \textbf{\num{96.21}} \\
     & \textbf{de-DE} & \textbf{\num{88.15}} & \num{86.04} & \num{86.63} & \num{91.05} & \num{89.47} & \textbf{\num{94.62}} \\
     & \textbf{es-US} & \num{104.27} & \textbf{\num{104.64}} & \num{103.65} & \num{102.24} & \num{101.83} & \textbf{\num{104.22}} \\ \bottomrule
    \end{tabular}
\end{table}

\vspace{-0.25cm}
Interestingly, results clearly indicate that for \textit{unified} features, bigger phoneme embedding sizes are consistently preferred over smaller ones for both naturalness and accent for all LVs. Such results are statistically significant with all p-value $< 0.005$. This seems to suggest that for \textit{unified} features, more phonetic information is captured in the phoneme embedding vectors, making higher dimensional vectors beneficial. 
In contrast, for \textit{separate} features, no particular patterns were detected, which lead us to conclude that $s=64$ is sufficient in this case. We hypothesise that the difference in results is explained by two factors. One is the increased total number of phonemes in the \textit{separate} features, which by itself increases the capacity of the model. The other is the reduced amount of times a phoneme is observed  when switching from \textit{separate} to \textit{unified} features. This allows each embedding to learn less information.
\vspace{-0.25cm}
\begin{table}[th]
    \centering
    \caption{Mean MUSHRA scores relative to recordings for Naturalness (N) and Accent (A), comparing \textit{separate} and \textit{unified} features at lower phoneme embedding sizes, across all evaluated LVs.}
    \label{tab:ablation-lower-comp}
    \begin{tabular}{@{}ll|ll|ll@{}}
    \toprule
     &  & \multicolumn{2}{c|}{\textbf{\num{64}}} & \multicolumn{2}{c}{\textbf{\num{128}}} \\
     &  & \multicolumn{1}{c}{Separate} & \multicolumn{1}{c|}{Unified} & \multicolumn{1}{c}{Separate} & \multicolumn{1}{c}{Unified} \\  \midrule
    \multirow{4}{*}{\textbf{N}} & \textbf{en-GB} & \num{77.03} & \textbf{\num{77.50}} & \textbf{\num{80.44}} & \num{76.52} \\
     & \textbf{fr-CA} & \textbf{\num{90.12}} & \num{87.88} & \num{90.11} & \textbf{\num{93.20}} \\
     & \textbf{de-DE} & \num{81.31} & \textbf{\num{84.48}} & \textbf{\num{80.17}} & \num{79.54} \\
     & \textbf{es-US} & \textbf{\num{92.89}} & \num{89.92} & \textbf{\num{100.40}} & \num{100.13} \\ \midrule 
    \multirow{4}{*}{\textbf{A}} & \textbf{en-GB} & \num{90.05} & \textbf{\num{90.16}} & \textbf{\num{89.02}} & \num{87.96} \\
     & \textbf{fr-CA} & \textbf{\num{91.70}} & \num{89.60} & \num{87.26} & \textbf{\num{90.49}} \\
     & \textbf{de-DE} & \num{87.99} & \textbf{\num{88.45}} & \textbf{\num{88.12}} & \num{87.29} \\
     & \textbf{es-US} & \num{108.97} & \textbf{\num{109.45}} & \num{107.30} & \textbf{\num{109.45}} \\ \bottomrule
    \end{tabular}
\end{table}
\vspace{-0.25cm}
The direct comparison between \textit{separate} and \textit{unified} features (Table \ref{tab:ablation-lower-comp}) confirms instead that the conclusions in Section \ref{subsec:phonetic-representations-comparison} are affected by the phoneme embedding size. In particular, we find that the differences between \textit{separate} and \textit{unified} features are less marked at lower embedding sizes, where the clear signal from Section \ref{subsec:phonetic-representations-comparison} disappears. Note in fact that the differences in relative mean scores observed between the two feature types are not statistically significant, for both naturalness and accent. 
In summary, the benefits of \textit{unified} representations start to manifest themselves only above a certain threshold of phoneme embedding size, while the quality of \textit{separate} representation models is not affected by a variation in phoneme embedding size.

\section{Conclusion}
\label{sec:conclusions}
In this work, we showed that a \textit{unified} phonetic representation can substantially improve the accent and naturalness of a Polyglot TTS model. Despite the phonetic representation being shared across all LVs, the model did not suffer from more accent quality degradation compared to the \textit{separate} representation model. Speaker similarity to the target speaker is also significantly improved compared to the model trained with \textit{separate} features, bringing it closer to that of single-speaker monolingual models. We also observed that the \textit{unified} representation model was more stable than the \textit{separate} representation model, likely due to a reduced number of tokens which allows for better generalisation over phonetic similarities between languages and their variants. Furthermore, we showed that the difference between \textit{unified} and \textit{separate} features only emerges at larger embedding sizes.
As an extension of this work, we could consider extending this work to languages from other language families. Moreover,we could also consider a comparison with a representation based on phonological features. 
These features may allow the model to learn about similarities between phonemes in different languages represented by different symbols.

\bibliographystyle{IEEEtran}

\bibliography{mybib}
\end{document}